\documentclass[graybox]{svmult}

\usepackage{mathptmx}       
\usepackage{helvet}         
\usepackage{courier}        
\usepackage{type1cm}        
\usepackage{makeidx}         
\usepackage{graphicx}        
\usepackage{multicol}        
\usepackage[bottom]{footmisc}

\usepackage[latin9]{inputenc}

\makeindex             

\begin{document}

\title*{When is a bottleneck a bottleneck?}
\author{Andreas Schadschneider, Johannes Schmidt and Vladislav Popkov}
\institute{Andreas Schadschneider \and Johannes Schmidt \and Vladislav Popkov
\at Universit\"at zu K\"oln, Institut 
f\"ur Theoretische Physik, 50937 K\"oln, Germany, 
\email{as@thp.uni-koeln.de, schmidt@thp.uni-koeln.de, vladipopkov@gmail.com}
}
%
%
\maketitle


\abstract{Bottlenecks, i.e. local reductions of capacity, are one
of the most relevant scenarios of traffic systems. The asymmetric 
simple exclusion process (ASEP) with a defect 
is a minimal model for such a bottleneck scenario. One crucial
question is "What is the critical strength of the defect that is
required to create global effects, i.e.  traffic jams localized at the
defect position". Intuitively one would expect that already an
arbitrarily small bottleneck strength leads to global effects in the
system, e.g. a reduction of the maximal current. Therefore it came as a
surprise when, based on computer simulations, it was claimed that the
reaction of the system depends in non-continuous way on the defect
strength and weak defects do not have a global influence on the
system. Here we reconcile intuition and simulations by
showing that indeed the critical defect strength is zero. We discuss
the implications for the analysis of empirical and numerical data.  
}


\section{Introduction}
\label{sec:intro}

One of the most important scenarios in any traffic system are
bottlenecks, i.e. (local) flow limitations. Typical examples are a
reduction in the number of lanes on a highway, local speed limits or
narrowing corridors or exits in pedestrian dynamics.  The
identification of bottlenecks gives important information about the
performance of the system. E.g. in evacuations, egress times are
usually strongly determined by the relevant bottlenecks. Therefore a
proper understanding of bottlenecks and their influence on properties
like the flow is highly relevant.

\begin{figure}[b]
\sidecaption[b]
\includegraphics[width=.5\textwidth]{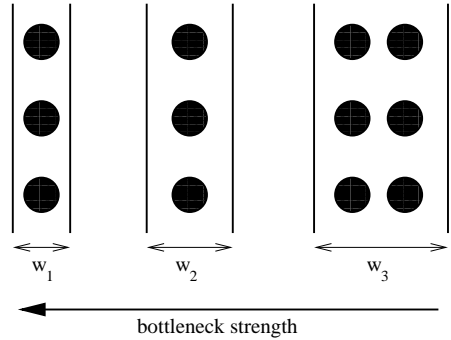}
%
%
\caption{Three corridors of different widths $w_j$. The bottleneck
  strength is inversely proportional to $w_j$. Lane formation leads to
  a non-continuous dependence of the current on the bottleneck
  strength.
}
\label{fig:zipper}       
\end{figure}

One of the most natural questions is "When does a bottleneck lead to a
traffic jam?" Does any bottleneck immediately lead to jam formation or
is there a minimal bottleneck strength required?  Intuitively one
would say that even a small bottleneck strength leads to
macroscropically observable effects, like a reduction of the maximal
current or jams.  However, other scenarios have been considered as
well and have even been part of legal guidelines. One prime example in
pedestrian dynamics is the dependence of the current on the width of a
corridor \cite{corridor2,corridor}.  Originally it was believed that
the current increases stepwise, i.e.  non-continuously, with increasing
bottleneck width. This increase was assumed to happen when the
corridor width allows an additional lane of pedestrians to be formed
(Fig.~\ref{fig:zipper}).
Taking the corridor width as measure for the bottleneck strength
(rather its inverse) this implies that an increasing bottleneck
strength not necessarily leads to smaller current values or jam
formation. In the meantime we know that this scenario is not correct
and the current increases linearly with the width \cite{corridor2}.
However it is still possible that there are situations where lane
formation is relevant and this scenario is more adequate, e.g. in
colloidal systems \cite{colloids}.

In the following we will take a theoretical physics point of view by
considering a minimal model for bottlenecks. Experience shows that the
results capture the generic nature of bottleneck transitions.

%


\section{Bottlenecks in the ASEP}
\label{sec:asep}



\begin{figure}[h]
\includegraphics[scale=.40]{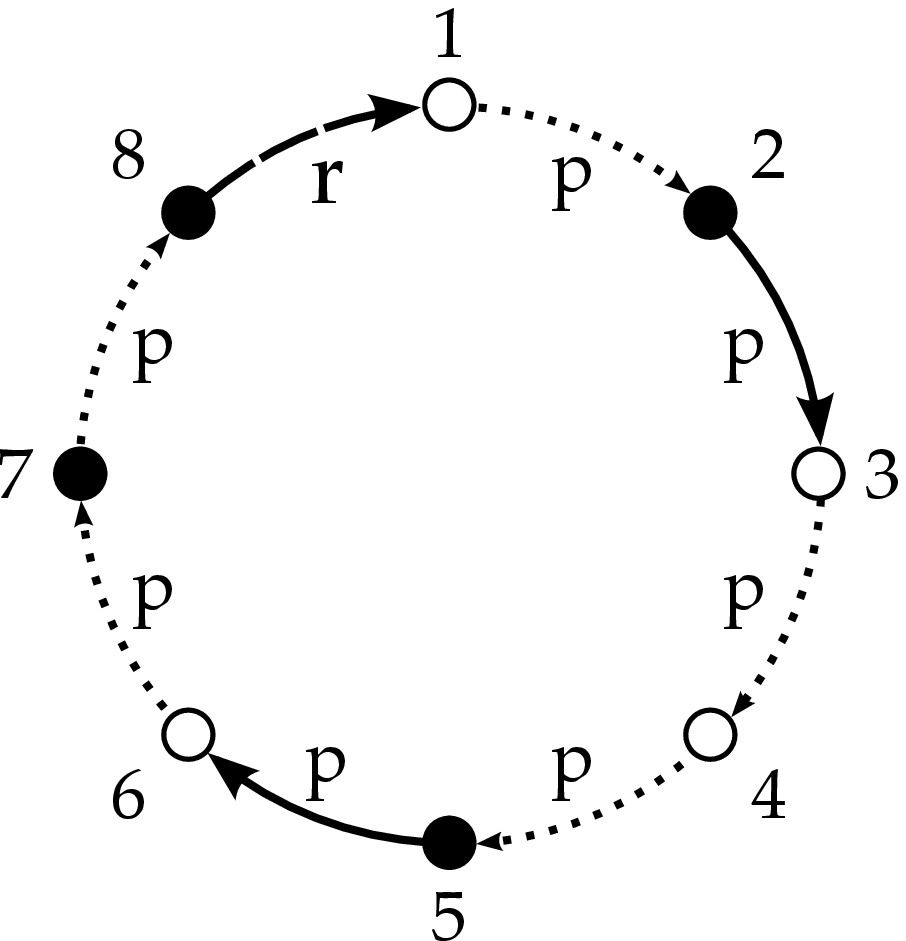}
\qquad
\includegraphics[scale=.42]{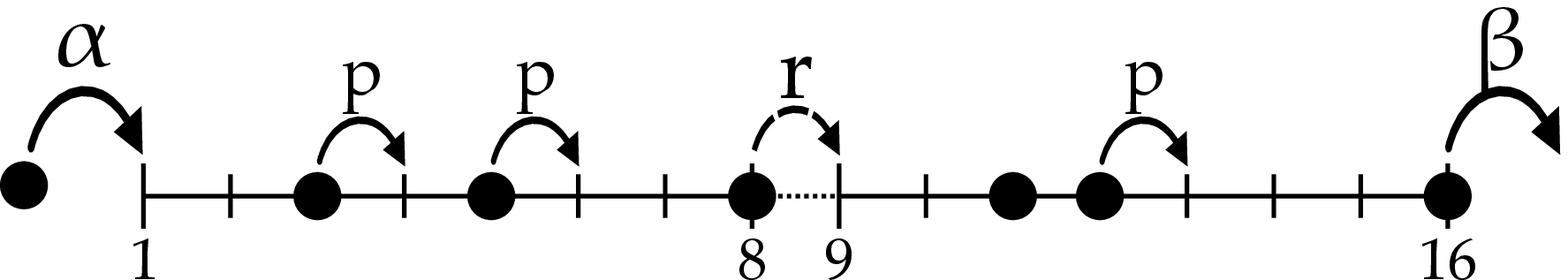}
%
%
\caption{ASEP with a defect (slow bond) where the hopping probability 
is $r<p$. $r=p$ corresponds to the homogeneous case.
Left: Periodic boundary conditions with $N=8$ sites, the
slow bond is between sites 8 and 1.
Right: Open boundary conditions with $N=16$ sites, the slow bond 
is between sites 8 and 9. 
}
\label{fig:ASEP}       
\end{figure}

The Asymmetric Simple Exclusion Process (ASEP) is a paradigmatic model
of of nonequilibrium physics (for reviews, see e.g.
\cite{bib:L,bib:D,bib:Schue,bib:BE,bib:SCN}) and arguably the simplest
model that captures essential features of traffic systems, i.e.
directed motion, volume exclusion and stochastic dynamics.  It
describes interacting (biased) random walks on a discrete lattice of
$N$ sites, where an exclusion rule forbids occupation of a site by
more than one particle. A particle at site $j$ moves to site $j+1$
with rate $p$ if site $j+1$ is not occupied by another particle
(Fig.~\ref{fig:ASEP}). In the following we will mainly use a
random-sequential update.  If sites are updated synchronously
(parallel update) the model is the $v_{\rm max}=1$ limit of the
Nagel-Schreckenberg model \cite{bib:SCN,NaSch}.  Many exact results
are known for the homogeneous case of the ASEP, e.g. the fundamental
diagram and the phase diagram in case of open boundary conditions
\cite{bib:L,bib:D,bib:Schue,bib:BE,bib:SCN}.

\begin{figure}[t]
\sidecaption[c]
\includegraphics[width=.55\textwidth]{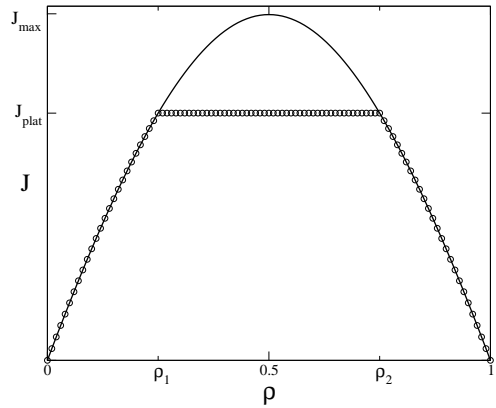}
%
%
\caption{Fundamental diagram of the ASEP with a defect $r$ (circles). The 
full line is the fundamental diagram of the homogeneous system without
defect. The current $J(r)$ is independent of the global density $\rho$
for $\rho_1<\rho<\rho_2$. The plateau value $J_{\rm plat}$ in
this region is smaller than the maximal flow $J_{\rm max}$ in the
homogeneous system.
}
\label{fig:defectFD}       
\end{figure}

A simple but generic model for a bottleneck is obtained by replacing
one of the hopping probabilities $p$ by a defect, or slow bond, with
hopping probability $r<p$ (Fig.~\ref{fig:ASEP}). Many properties of
this defect system have been obtained in a seminal paper by Janowsky
and Lebowitz \cite{JanLeb1}. They have shown that the shape of the
fundamental diagram can be understood by a simple mean-field theory.
In the stationary state the current can be obtained by matching the
current $J_{\rm hom}$ in the homogeneous system with the current
$J_{\rm def}$ at the defect.  Neglecting correlations at the defect
site one finds that the defect has no influence on the system for low
densities $\rho < \rho_1$ and large densities $\rho> \rho_2$
\footnote{For the ASEP, due to particle-hole symmetry, $\rho_1 =
  1-\rho_2$.}. The density remains uniform throughout the whole system
and the current is identical to that of the homogeneous system
(Fig.~\ref{fig:defectFD}).

\begin{figure}[h]
\includegraphics[scale=.48]{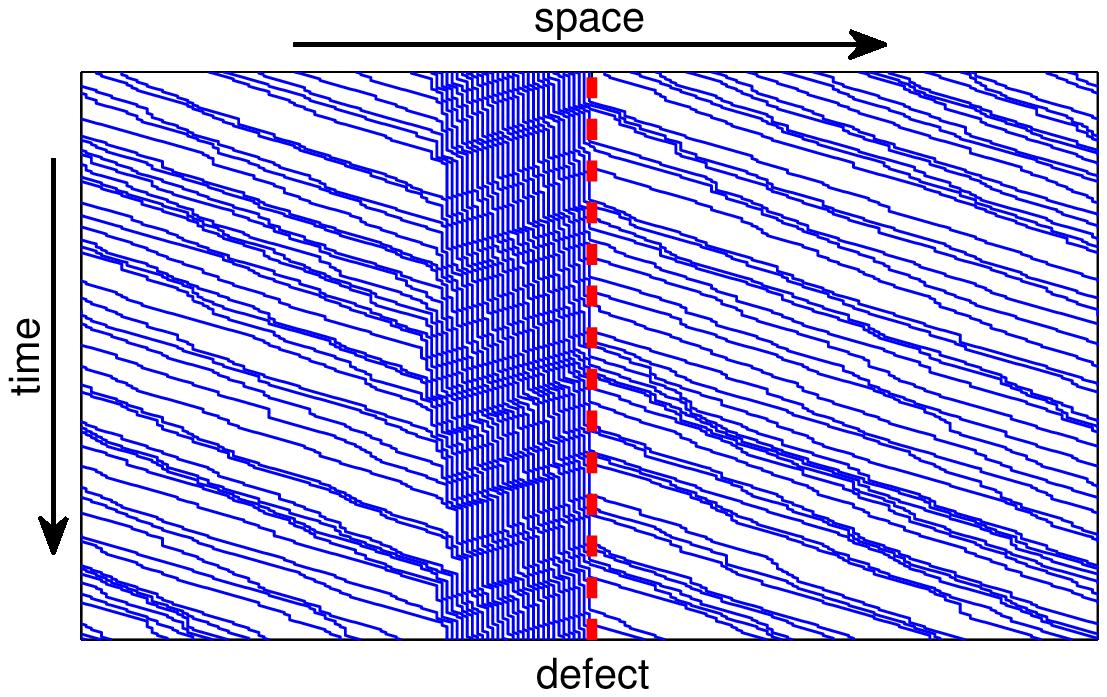}
\quad
\includegraphics[scale=.48]{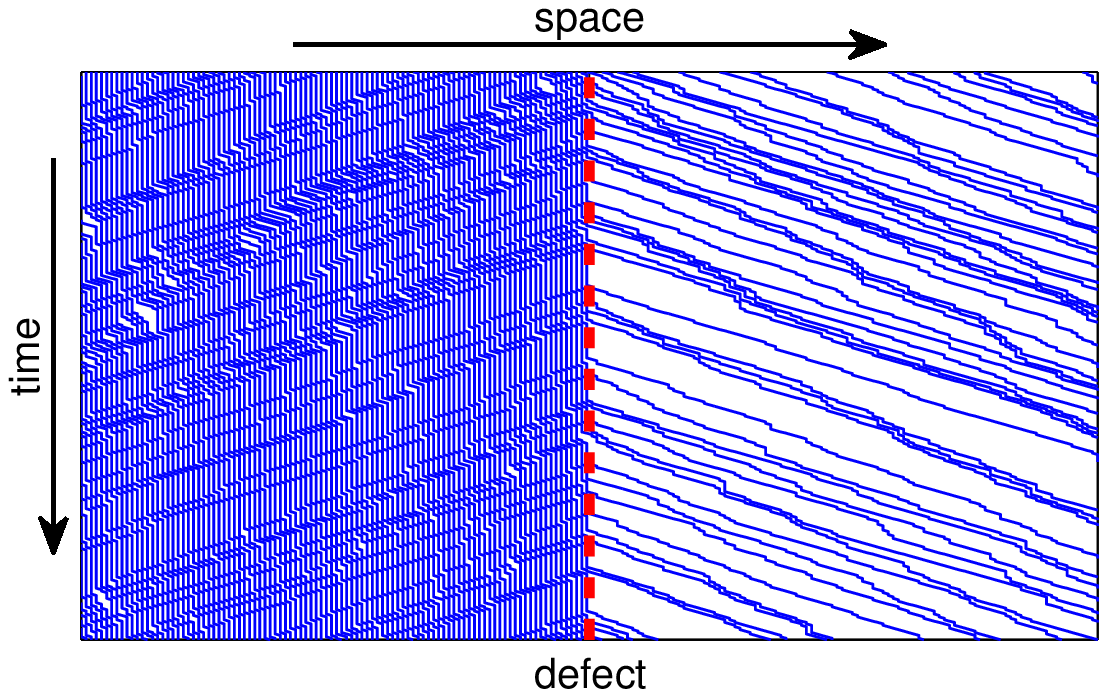}
%
%
\caption{Phase separation in the plateau regime. Left: Periodic boundary
conditions. Right: Open boundary conditions.
}
\label{fig:phasesepa}       
\end{figure}

For densities $\rho_1 < \rho < \rho_2$, on the other hand, the
fundamental diagram exhibits a plateau where the current is
independent of the density (Fig.~\ref{fig:defectFD}). The plateau
value $J_{\rm plat}$ corresponds to the maximal current that is
supported by the defect. In this density regime the stationary state
is no longer characterized by a uniform density. Instead phase
separation into a high and a low density region is observed. The high
density region corresponds to a jam that is formed at the defect
position (Fig.~\ref{fig:phasesepa}). For periodic boundary conditions
the length of jam shows characteristic fluctuations 
(Fig.~\ref{fig:phasesepa}, left) \cite{JanLeb1}.

For the ASEP with periodic boundary conditions, random-sequential
update and a defect $r$ mean-field theory makes quantitative
predictions for the phase separated regime \cite{JanLeb1}. The
value of the current in the plateau region is given by
\begin{equation}
J_{\rm plat} = \frac{pr}{(p+r)^2} 
\label{leboflux}
\end{equation}
and the densities in the low and high density region by
\begin{equation}
\rho_{\ell}  = \frac{r}{p+r}\,.
\qquad {\rm and} \qquad 
\rho_h  = \frac{p}{p+r} 
\label{density}
\end{equation}
The critical densities $\rho_1$, $\rho_2$ which determine the 
plateau regime $\rho_1<\rho<\rho_2$ are simply
\begin{equation}
\rho_1 = \rho_{\ell}
\qquad {\rm and} \qquad 
\rho_2 = \rho_h\,.
\label{density2}
\end{equation}
The mean-field results are supported by systematic series expansions
\cite{expansions}.

\begin{figure}[h]
\sidecaption[b]
\includegraphics[width=.55\textwidth]{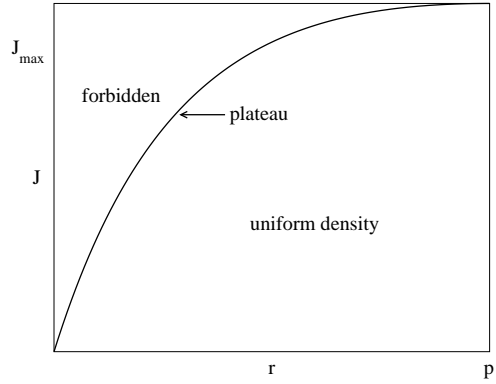}\\
\caption{Phase diagram of the ASEP with defect according to
\cite{JanLeb1}. The full line shows the current at the plateau
as function of the defect hopping rate~$r$. Any $r<p$ leads
to a reduction of the maximal current compared to that of
the homogeneous system~$J_{\rm max}$. In the phase of uniform
density the defect has only local effects.
}
\label{fig:phase1}       
\end{figure}

Fig.~\ref{fig:phase1} shows the resulting phase diagram. For any
defect $r< p$ only currents up to the plateau value $J_{\rm plat}$
can be realized in the system which then phase separates into
a high density region pinned at the defect and a low density
regime. For currents $J<J_{\rm plat}$ the density is uniform.
The important point is that $J_{\rm plat}< J_{\rm max}$ for
any $r<p$ where $J_{\rm max}$ is the maximal current in the
homogeneous system. In other words: any bottleneck leads to
a reduction of the current and a phase separated state (at
intermediate densities).

\section{What is the critical bottleneck strength?}
\label{sec:3}

\begin{figure}[h]
\sidecaption[b]
\includegraphics[width=.55\textwidth]{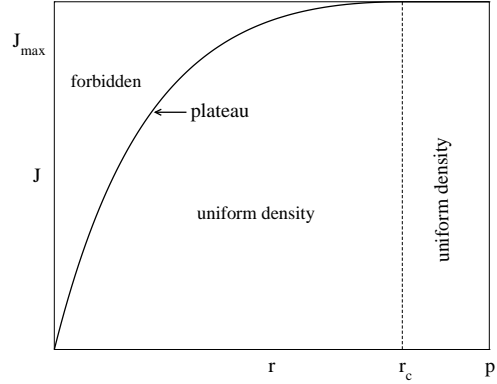}
\caption{Phase diagram of ASEP with defect according to \cite{Ha}.
Defects with $r_c < r \leq p$ have no influence on the current $J$.
}
\label{fig:phase2}       
\end{figure}

Mean-field theory predicts that any bottleneck $r<p$ leads to the
formation of a plateau in the fundamental diagram and the 
associated phase-separated state \cite{JanLeb1}. Defining the
bottleneck strength by 
\begin{equation}
\Delta p = \frac{p-r}{p}
\end{equation}
this implies that the critical bottleneck strength $(\Delta p)_c$ at
which the defect has global influence on the system (e.g. its current
or the density) is predicted to be
\begin{equation}
\left(\Delta p\right)_c = 0\,,\qquad {\rm i.e.}\quad r_c = p\,.
\end{equation}
As mentioned in the Introduction this is what is intuitively expected.
Therefore it came as quite a surprise when it was claimed \cite{Ha},
based on extensive computer simulations, that $r_c \approx 0.8$, i.e.
\begin{equation}
\left(\Delta p\right)_c^{\rm (Ha)} \approx 0.2\,.
\end{equation}
The corresponding phase diagram is shown in Fig.~\ref{fig:phase2}.
In contrast to Fig.~\ref{fig:phase1}, for defects $r>r_c$ all
currents up to $J_{\rm max}$ can be realized and there is no
phase separation at any density for weak defects! In this case
the bottleneck has only {\em local} effects which can be observed
near the defect, but not in the whole system.

Due to this apparent contradiction with expectations we have revisited
the ASEP defect problem in \cite{Schmidt} based on highly accurate
Monte Carlo simulations. Similar to \cite{Ha} we have simulated the
ASEP with open boundary conditions, random-sequential dynamics (with
$p=1$) and a defect in the middle of the system (Fig.~\ref{fig:ASEP}).
However, choosing $\alpha=\beta=\frac{1}{2}$ as in \cite{Ha},
corresponds exactly to the phase boundary of the high, low and maximal
current phase \cite{bib:D,bib:Schue,bib:SCN}.  Fluctuations in
finite-size systems will systematically underestimate the defect
current $J\left(r\right)$ \cite{Schmidt}.  We have therefore choosen
$\alpha=\beta=1$ well inside the maximal current phase which allows to
obtain a much better statistics.

To determine rather subtle bottleneck effects, very good statistics and
advanced Monte Carlo techniques are required. To minimizes errors 
induced by pseudo-random number generators we have used
the  Mersenne Twister \cite{Schmidt}.  

Measurements of bottleneck effects for small defect strengths are
easily hidden by fluctuations. Instead of using independent
measurements for each defect strength $r$ the systems are evolved
in parallel, i.e. with the same protocol and the same set of random
numbers, which leads to a strong suppression of fluctuations~\cite{Schmidt}.

\begin{figure}[h]
\sidecaption[b]
\includegraphics[width=.6\textwidth]{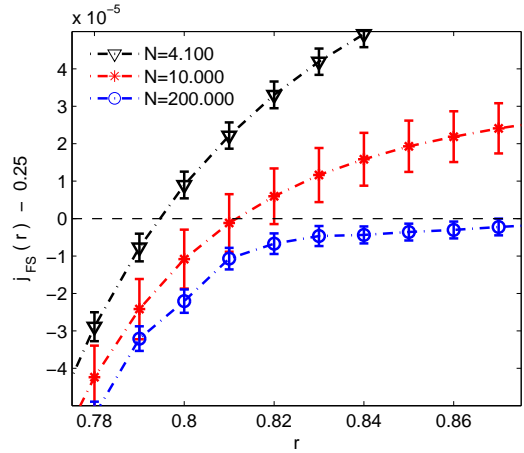}
\caption{Finite-size corrections to the current. The exactly known
current in the infinite homogeneous system is $J(N=\infty,r=1)=1/4$.}
\label{fig:fsc}       
\end{figure}

In order to minimize finite-size corrections, system lengths of up to
$N=200.000$ were considered (Fig.~\ref{fig:fsc}) which is two orders of
magnitude larger than the systems considered in \cite{Ha}.

To estimate the global effects of the defect we first considered the
finite-size current $J(N,r)$ through a system of length $N$ and with a
defect $r$. Due to the fact that finite size corrections lead to an
enhanced current, i.e.  $J\left(r,N\right)>J\left(r,N=\infty\right)$,
one finds a lower bound for the critical hopping rate by satisfying
$J\left(N,r_{c}\right)-J\left(N=\infty,r=1\right)<0$.  However, in
this way we only could derive a lower bound $r_{c}\geq 0.86$ for the
critical hopping rate (Fig.~\ref{fig:fsc}).  Assuming the existence of
an essential singularity at $r_{c}=1$, $i.e.$
$j\left(1\right)-j\left(r\right)\sim\exp\left(-a/\left(
    1-r\right)\right)$ \cite{expansions}, further improvement of the
lower bound for the critical defect $r_c$ by increasing the system
length is a hopeless enterprise: e.g. a numerical proof of $r_{c}>0.9$,
$r_{c}>0.95$, $r_{c}>0.99$ would require $N>10^{10}$, $N>10^{22}$,
$N>10^{147}$, respectively.

\begin{figure}[h]
\sidecaption[b]
\includegraphics[width=.6\textwidth]{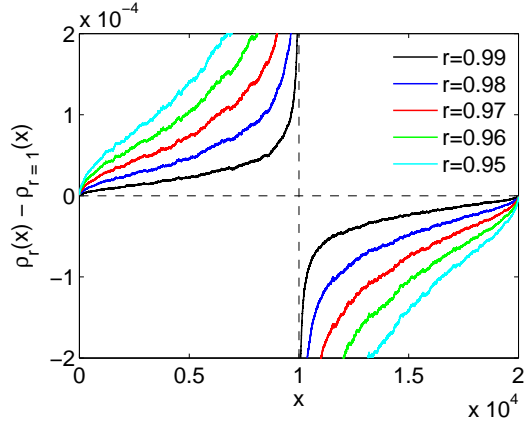}
\caption{Arbitrary defects $r$ have a non-local effect on 
the density profile. The figure shows the difference between
the density profile $\rho_r(x)$ with and that without defect
$\rho_{r=1}(x)$ where $x=j/N$ is the rescaled position.}
\label{fig:densprof}       
\end{figure}

A much better quantity to determine the global influence of the defect
(see e.g. Fig.~\ref{fig:phasesepa}, right) is the density profile or
rather the difference between the density profile of the defect system
with a corresponding homogeneous system (Fig.~\ref{fig:densprof}).
Using the approach of parallel evolving systems we could clearly show
a nonlocal influence on the density profile for defect strengths up to
$r=0.99$ (Fig.~\ref{fig:densprof}).  This strongly supports the
mean-field prediction $r_{c}=1$.

\section{Discussion and relevance for empirical results}

Despite its relevance for applications some fundamental aspects
of bottlenecks are not fully understood. Even for a minimal model
like the ASEP with a defect the influence of weak bottlenecks is rather
subtle and can be easily lost in fluctuations.

We have shown how to reconcile computer simulations with the intuition
that even small defects have a global influence on the system.
These effects are not easily seen in a reduction of the current which
presumably shows a non-analytic dependence on the bottleneck strength.
Bottlenecks are better identified by their effects on the density
profile which spreads throughout the whole system..

Based on a careful statistical analysis of Monte Carlo simulations
we have found strong evidence that an arbitrarily weak defect 
$\Delta p\to 0$ in the ASEP has a global influence on 
the system. Meanwhile a mathematical proof of $\left(\Delta p\right)_c=0$
has been announced in \cite{Sidoravicius2015}.

These results are believed to be generic for bottleneck systems.
As a consequence the identification of weak bottlenecks in noisy
empirical data is extremely difficult. Even for computer simulations
very good statistics is required. Since the effect on the current
is rather small, the density profile might be
a better indicator for the presence of weak bottlenecks.



\begin{acknowledgement}
We dedicate this contribution to the memory of our friend and
colleague Matthias Craesmeyer. Financial support by the DFG under 
grant SCHA 636/8-1 is gratefully acknowdledged.
\end{acknowledgement}
%



\bibliographystyle{spmpsci.bst}

\end{document}